\documentclass[11pt]{article}
\newcommand{\nc}{\newcommand}
\nc{\bib}{\bibitem}
\nc{\al}{\alpha}
\nc{\g}{\gamma}
\nc{\G}{\Gamma}
\nc{\D}{\Delta}
\nc{\eps}{\epsilon}
\nc{\la}{\lambda}
\nc{\La}{\Lambda}
\nc{\var}{\varphi}
\nc{\pa}{\partial}
\nc{\nn}{\nonumber \\ }
\nc{\hf}{\frac{1}{2}}
\nc{\dz}{\frac{dz}{2\pi i}}
\nc{\bin}[2]{\left (\begin{array}{c} {#1}\\ {#2} \end{array}\right )}
\nc{\ben}{\begin{equation}}
\nc{\een}{\end{equation}}
\nc{\bea}{\begin{eqnarray}}
\nc{\eea}{\end{eqnarray}}
\nc{\bra}[1]{\langle {#1}|}
\nc{\ket}[1]{|{#1}\rangle}
\newcommand{\Z}{\mbox{$Z\hspace{-2mm}Z$}}
\nc{\C}{\mbox{\hspace{1.24mm}\rule{0.2mm}{2.5mm}\hspace{-2.7mm} C}}
\nc{\Nat}{\mbox{\hspace{.04mm}\rule{0.2mm}{2.8mm}\hspace{-1.5mm} N}}
\newcommand{\R}{\mbox{\hspace{.04mm}\rule{0.2mm}{2.8mm}\hspace{-1.5mm} R}}

\nc{\NP}[1]{Nucl.\ Phys.\ {\bf #1}}
\nc{\PL}[1]{Phys.\ Lett.\ {\bf #1}}
\nc{\CMP}[1]{Commun.\ Math.\ Phys.\ {\bf #1}}
\nc{\PR}[1]{Phys.\ Rev.\ {\bf #1}}
\nc{\PRL}[1]{Phys.\ Rev.\ Lett.\ {\bf #1}}
\nc{\PTP}[1]{Prog.\ Theor.\ Phys.\ {\bf #1}}
\nc{\PTPS}[1]{Prog.\ Theor.\ Phys.\ Suppl.\ {\bf #1}}
\nc{\MPL}[1]{Mod.\ Phys.\ Lett.\ {\bf #1}}
\nc{\IJMP}[1]{Int.\ Jour.\ Mod.\ Phys.\ {\bf #1}}
\nc{\IM}[1]{Invent.\ Math.\ {\bf #1}}
\nc{\SJNP}[1]{Sov. J. Nucl. Phys.\ {\bf #1}}
\nc{\JHEP}[1]{J.\ High\ Energy Phys.\ {\bf #1}}

\def\vvdots{\mathinner{\mkern1mu\raise1pt\vbox{\kern7pt\hbox{.}}\mkern2mu
 \raise4pt\hbox{.}\mkern2mu\raise7pt\hbox{.}\mkern1mu}}

\def\max{{\rm max}}
\def\min{{\rm min}}

\begin{document}

\topmargin -5mm
\oddsidemargin 5mm

\begin{titlepage}
\setcounter{page}{0}

\vspace{8mm}
\begin{center}
{\huge Higher-genus $su(N)$ fusion multiplicities}\\[.4cm]
{\huge as polytope volumes}

\vspace{15mm}
{\large G. Flynn}$^{\star1}$, 
{\large J. Rasmussen}$^{\dagger}$,
{\large M. Tahi\'c}$^{\star}$\footnote{GF and MT were supported in part by 
NSERC Undergraduate Student Research Awards.} 
{\large and M.A. Walton}$^{\star}$
\\[.2cm]
$^{\star}${\em Physics Department, University of Lethbridge,
Lethbridge, Alberta, Canada T1K 3M4}\\
$^{\dagger}${\em CRM, Universit\'e de Montr\'eal, Case postale 6128, 
succursale centre-ville, Montr\'eal, Qu\'ebec, Canada H3C 3J7}
\\[.3cm]
{E-mail: flynngt@uleth.ca, rasmusse@crm.umontreal.ca, 
tahimk@uleth.ca, walton@uleth.ca}

\end{center}

\vspace{10mm}
\centerline{{\bf{Abstract}}}
\vskip.4cm
\noindent
We show how higher-genus $su(N)$ fusion multiplicities may be computed
as the discretized volumes of certain polytopes. The method is 
illustrated by explicit analyses of some $su(3)$ and $su(4)$
fusions, but applies to all higher-point and higher-genus $su(N)$ fusions. 
It is based on an extension of the realm of Berenstein-Zelevinsky
triangles by including so-called gluing and loop-gluing diagrams.
The identification of the loop-gluing diagrams is our main
new result, since they enable us to characterize higher-genus fusions
in terms of polytopes. Also, the genus-2 0-point $su(3)$ fusion
multiplicity is found to be a simple binomial coefficient
in the affine level.

\end{titlepage}
\newpage
\renewcommand{\thefootnote}{\arabic{footnote}}
\setcounter{footnote}{0}

\section{Introduction}

Recently, methods have been developed for computing $su(N)$
tensor product \cite{RW1,RW2} and fusion \cite{RW3,RW4}
multiplicities based on a generalization of the Berenstein-Zelevinsky
(BZ) triangles \cite{BZ}. The idea is to associate a convex polytope to
a multiplet of integrable highest weights $(\la,...,\sigma)$.
The discretized volume of the polytope is the (tensor product or) fusion
multiplicity associated to the coupling of the weights
$(\la,...,\sigma)$ to the singlet. 

Ref. \cite{RW1} describes ordinary
three-point tensor products, while \cite{RW2} extends the results
to higher-point couplings. The extension is obtained by introducing a
{\em gluing} of BZ triangles, whereby the triangular configurations
are replaced by multi-sided configurations or diagrams.
   
The dependence on the affine level in fusion may be implemented 
by associating threshold levels to the tensor product couplings
\cite{CMW}. Using that idea in the framework of BZ triangles,
allows one to characterize also fusion multiplicities by 
polytopes. The extra input is an assignment of threshold
levels to the BZ triangles. That is trivial for $su(2)$,
straightforward for $su(3)$ \cite{KMSW}, somewhat complicated for $su(4)$
\cite{BKMW}, but believed to be possible for all $su(N)$.
Polytope characterizations of fusion multiplicities have been studied 
in \cite{RW3,RW4,osp}: ${\cal N}$-point $su(2)$ and $osp(1|2)$ fusions 
are treated in \cite{RW2} and \cite{osp}, respectively, 
while \cite{RW4} discusses three-point $su(3)$ and $su(4)$ 
fusions. One objective of the present work is the extension of the latter
results to higher-point fusions.

Higher-genus fusions may also be characterized by polytopes.
The appearance of loops forces us to introduce a new class of diagrams. 
We shall call them {\em loop-gluing diagrams},
or for short, {\em loop gluings}. In the case of $su(2)$, they were
introduced in \cite{RW3}, leading to a characterization of all 
higher-genus ${\cal N}$-point $su(2)$ fusions by convex polytopes.
This was extended to $osp(1|2)$ in \cite{osp}, and
the second objective of the present work is the extension of it
to $su(N)$. The main results are the identification of the
$su(N)$ loop gluings (illustrated for $su(4)$ in (\ref{gl4})), 
and the explicit characterization of 
higher-genus ${\cal N}$-point $su(3)$ and $su(4)$ fusion multiplicities
as discretized volumes of certain polytopes.

The characterization of fusion multiplicities as the discretized volume
of polytopes, makes manifest that the multiplicities are non-negative 
integers.
That is a priori not clear when examining the Verlinde formula \cite{Ver}.
Furthermore, the geometrical interpretation offers a better understanding
of the underlying symmetries and properties of the multiplicities and
their level-dependence. As an example, in \cite{RW4} it was
conjectured that for a fixed triplet of
$su(N)$ weights $(\la,\mu,\nu)$, the threshold
multiplicity has at most one local maximum as a discrete
function of the threshold level. We recall that the threshold multiplicity
\cite{irvine} is the number of different couplings of the weights 
$(\la,\mu,\nu)$ with the same threshold level.

As a computational advantage of our description we mention that 
it results in fast computer programs, when the discretized volumes of the 
polytopes are measured in terms of multiple-sum formulas. The latter also 
provide very explicit formulas for the multiplicities, as opposed to the 
well-known combinatorial ones as the Littlewood-Richardson rule for tensor 
products, for example. More conjecturally, the geometrical interpretation may 
help toward an extended Littlewood-Richardson rule for fusion.

Our results are of a high technical complexity. For the benefit of this
presentation, we thus focus on examples and only allude to the general
case. It is straightforward to describe, though. After a brief discussion
of BZ triangles and the method of gluing, we introduce the loop-gluing
diagrams in Section 2. An application is considered in Section 3
where we examine the genus-1 ${\cal N}$-point $su(3)$ fusion
multiplicities for generic non-negative integer level.
The genus-2 0-point $su(3)$ fusion
multiplicity is worked out explicitly and found to be a simple
binomial coefficient in the 
affine level. This is believed to be the first concise result on 
fusion multiplicities for rank and genus both higher than one.
In Section 4 we describe the extension to higher genus and
higher-rank $su(N)$. We pay particular attention to genus-1 
0-point fusions for higher rank. Section 5 contains some concluding remarks,
while Appendix A provides details on higher-genus $su(4)$
fusion multiplicities, with particular emphasis on the genus-1 1-point
and genus-2 0-point fusions.

\section{Triangles, gluings and loops}

  Let us briefly review some results of \cite{RW1,RW2,RW3,RW4}. We refer
  to those papers for details. 

  A BZ triangle is a triangular arrangement of non-negative integers
  subject to certain constraints: the outer constraints depending on 
the three 
  $su(N)$ weights $(\la,\mu,\nu)$, and the hexagon identities which are 
  consistency conditions. According to \cite{BZ}, the number of possible BZ 
  triangles determines the associated tensor product multiplicity 
  $T_{\la,\mu,\nu}$.

  By relaxing the constraint that all the $E_r=\frac{3}{2}r(r+1)$ integer 
  entries are non-negative, 
  one may express any such generalized BZ triangle 
  ${\cal T}$ as the linear combination \cite{RW1}
  \ben
   {\cal T}={\cal T}_0+\sum_{i,j\geq1}^{i+j=r}v_{i,j}{\cal V}_{i,j}
  \label{TTV}
  \een
  Here $r$ is the rank of $su(r+1)$, while $\{{\cal V}_{i,j}\}$ is the
  set of associated virtual triangles -- one for each of the $H_r=r(r-1)/2$ 
  hexagons. ${\cal T}_0$ is an initial (generalized BZ) triangle, and
  the linear coefficients $v_{i,j}$ are integers. A convenient choice of
  initial triangle may be found in \cite{RW1}.
  A simple choice of labeling of the virtual triangles follows from
  \ben
  \matrix{\star\cr
          \star~~\qquad \star\cr
   \star\qquad v_{r-1,1}\quad \star\cr
   \star\qquad\ \star\qquad\ \ \star\ \qquad \star\cr
   \star~\quad\ v_{r-2,1}\quad \star\quad\ v_{r-2,2}\quad~\star \cr
   \star\quad\ \ \star\qquad \ \star\quad~~\quad \star\qquad\ \star\quad\ \ 
    \star \cr
   \vvdots\ 
    \qquad\qquad\qquad\vdots\ \ \ \qquad\vdots\qquad\qquad\qquad\ddots\cr
   \star \quad\qquad\qquad\qquad\qquad\quad
     \qquad\qquad\qquad\qquad\qquad \star\cr
          \star~~\qquad \star \ \quad\qquad\qquad\qquad\qquad\quad
     \qquad\qquad\qquad \star~~\qquad \star\cr
   \star~\qquad v_{2,1}\quad ~~\star
     \qquad\qquad\qquad \dots \qquad\qquad\qquad
   \star~\quad v_{2,r-2}\quad ~~\star\cr
   \star~\quad\ \ \star~~\qquad \star \ \qquad \star 
      \qquad\qquad\qquad\qquad\qquad
   \star\qquad\ \star~~\qquad \star\ \ \quad ~\star\cr
   \star\qquad v_{1,1} \ \quad \star\qquad v_{1,2}\ \ \quad \star 
       \qquad\quad\qquad\qquad\
   \star\quad\ \ v_{1,r-2}\quad \star\quad\ \ v_{1,r-1}\quad \star \cr
   \star\qquad \star\ \ \quad \star~~\qquad \star\qquad \ \star\qquad~ 
    \star  
       \quad\ \dots \quad\
   \star\qquad~ \star\qquad \star~~\qquad \star\qquad \star\qquad 
    \star \cr\cr}
  \label{hexnot}
  \een
  A $\star$ denotes an unspecified entry, and we see the hexagon structures
  surrounding each of the linear coefficients.
  The virtual triangles correspond to the simple distribution
  \ben
   \matrix{\matrix{1\cr 1\quad\bar1~~\quad\bar1\quad 1\cr
    \bar1~\quad\quad ~~\bar1\cr
    1~\quad\bar1~\quad\bar1~\quad 1\cr
   1\cr}}
  \label{virt}
  \een
  of plus and minus ones ($\bar1\equiv-1$) to any given hexagon. All other 
  entries are zero. Re-imposing the constraint that all entries of ${\cal T}$ 
  (\ref{TTV}) must be non-negative, results in a set of inequalities in 
  $\{v_{i,j}\}$.  They define a convex polytope, whose discretized volume is 
  $T_{\la,\mu,\nu}$ \cite{RW1}. A BZ triangle with all entries
  {\em non-negative} integers is called a true BZ triangle.

  Higher-point tensor products may be treated by gluing triangles together.
  The idea stems from the well-known decomposition
  \ben
   T_{\la,\mu,\nu,\sigma}=\sum_\tau T_{\la,\mu,\tau}T_{\nu,\sigma,\tau^+}
  \label{TTT}
  \een
  where $\tau^+$ is the weight conjugate to $\tau$. In terms of diagrams
  we have
  \ben
  \mbox
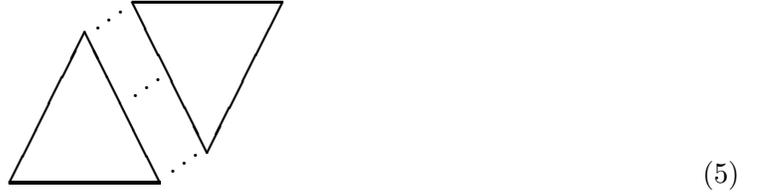
{
  \begin{picture}(100,80)(0,0)
  \unitlength=1cm
  \thicklines
   \put(0,0){\line(1,2){1}}
   \put(0,0){\line(1,0){2}}
   \put(2,0){\line(-1,2){1}}
   \put(1.1,2){$\vvdots$}
   \put(2.1,0.1){$\vvdots$}
   \put(1.6,1.1){$\vvdots$}
   \put(1.5,2.4){
  \begin{picture}(50,50)
   \put(0,0){\line(1,-2){1}}
   \put(0,0){\line(1,0){2}}
   \put(2,0){\line(-1,-2){1}}
  \end{picture}}
  \end{picture}
  }
  \label{four}
  \een
  where each of the triangles represent a BZ triangle. Along the glued
  interface, indicated by dotted lines,
  the weights must match. We recall that the three weights
  $(\la,\mu,\nu)$ of a coupling to the singlet are related as
  $\la+\mu-\nu^+=\sum_{i=1}^rn_i\al_i$ with $n_i\in\Z_\geq$.
  $\{\al_i\ |\ i=1,...,r\}$ is the set of simple roots.  
  Thus, we are led to introduce
  {\em gluing diagrams} or {\em gluing roots}, which correspond to combining 
  the two BZ triangles (as in (\ref{four})) associated to the couplings
  $(0,0,\al_i)$ and $(0,0,\al_i^+)$, respectively. 
  Let us illustrate the general construction \cite{RW2} by listing the two 
  gluing roots for $su(3)$:
  \ben
  \mbox{
  \begin{picture}(100,100)(125,0)
  \unitlength=1cm
  \thicklines
  \put(-1,1.5){${\cal G}_1\ \ =$}
  \put(0,0){0}
  \put(1,0){0}
  \put(2,0){$\bar1$}
  \put(3,0){1}
  \put(0.5,0.7){0}
  \put(2.5,0.7){1}
  \put(1,1.4){0}
  \put(2,1.4){$\bar1$}
  \put(1.5,2.1){0}
  \put(1.8,2.4){$\vvdots$}
  \put(2.3,1.7){$\vvdots$}
  \put(2.8,1){$\vvdots$}
  \put(3.3,0.3){$\vvdots$}
  \put(2.4,2.7){0}
  \put(3.4,2.7){0}
  \put(4.4,2.7){0}
  \put(5.4,2.7){0}
  \put(2.9,2.0){$\bar1$}
  \put(4.9,2){0}
  \put(3.4,1.3){1}
  \put(4.4,1.3){$\bar1$}
  \put(3.9,0.6){1}
  \put(7.2,0){\begin{picture}(100,100)
  \put(-1,1.5){${\cal G}_2\ \ =$}
  \put(0,0){0}
  \put(1,0){0}
  \put(2,0){0}
  \put(3,0){0}
  \put(0.5,0.7){0}
  \put(2.5,0.7){$\bar1$}
  \put(1,1.4){$\bar1$}
  \put(2,1.4){1}
  \put(1.5,2.1){1}
  \put(1.8,2.4){$\vvdots$}
  \put(2.3,1.7){$\vvdots$}
  \put(2.8,1){$\vvdots$}
  \put(3.3,0.3){$\vvdots$}
  \put(2.4,2.7){1}
  \put(3.4,2.7){$\bar1$}
  \put(4.4,2.7){0}
  \put(5.4,2.7){0}
  \put(2.9,2.0){1}
  \put(4.9,2){0}
  \put(3.4,1.3){$\bar1$}
  \put(4.4,1.3){0}
  \put(3.9,0.6){0}
  \end{picture}}
  \end{picture}
  }
  \label{GG}
  \een
  The extension to ${\cal N}$-point couplings, ${\cal N}>4$, is
  straightforward. We are thus extending triangles to ${\cal N}$-sided 
  diagrams ${\cal D}$, and (\ref{TTV}) is replaced by
  \ben
   {\cal D}={\cal D}_0+\sum_{a=1}^{{\cal N}-2}
    \sum_{i,j\geq1}^{i+j=r}v_{i,j}^{(a)}
    {\cal V}_{i,j}^{(a)}-\sum_{a=1}^{{\cal N}-3}\sum_{i=1}^rg_i^{(a)}
    {\cal G}_i^{(a)}
  \label{DVG2}
  \een
  The initial diagram ${\cal D}_0$ is easily described, see \cite{RW2}. The 
  label $a$ runs over the participating triangles or gluings, respectively,
  while the sign in front of the last term merely is for convenience.

  Fusion multiplicities are determined from the tensor product
  multiplicities and the associated multi-set of threshold levels $\{t\}$
  \cite{CMW}. In order to extend our polytope characterization of tensor 
  product multiplicities to fusion multiplicities, we need a prescription for 
  assigning a threshold level to a BZ triangle. That is trivial for
  $su(2)$, and was worked out for $su(3)$ in \cite{KMSW} and for
  $su(4)$ in \cite{BKMW}. The extension to higher rank is not known
  explicitly, but believed to exist. 

Assigning a threshold level $t$ to a triangle ${\cal T}$ amounts to expressing
  $t$ in terms of the entries of ${\cal T}$. In the known cases, $t$ is given 
  as a maximum over simple expressions in the entries. This leads 
  straightforwardly to a refinement of the polytope associated to the 
  underlying tensor product coupling, by introducing inequalities 
depending on 
  (and hence incorporating) the dependence on the affine level $k$. The 
  procedure will be illustrated explicitly below.

  To treat higher-genus fusion, we need to understand how loops appear at the 
  level of our diagrams. It is sufficient to focus on the ``self-gluing'' or 
  tadpole:
  \ben
  \mbox{
  \begin{picture}(100,40)(-25,-20)
  \unitlength=0.7cm
  \thicklines
  \put(1.15,0){\line(-2,1){1}}
  \put(1.15,0){\line(-2,-1){1}}
  \put(0.15,0.5){\line(0,-1){1}}
  \thinlines
  \put(1.15,0){\circle{1.4}}
  \put(-0.5,0){\line(1,0){0.95}}
  \end{picture}
  }
  \label{tad}
  \een
  The dual picture of ordinary (Feynman-like) graphs is shown in 
thinner lines.
  Let us consider $su(2)$ first, in which case the associated 
  {\em loop-gluing diagram} is
  \ben
  \mbox{
  \begin{picture}(100,30)(-25,-10)
  \unitlength=0.7cm
  \thicklines
  \put(0,0.5){0}
  \put(0,-0.5){0}
  \put(1,0){1}
  \end{picture}
  }
  \label{gl2}
  \een
  It is stressed that it differs radically from the $su(2)$ gluing root 
  \ben
  \mbox{
  \begin{picture}(100,65)(-45,-25)
  \unitlength=1cm
  \thicklines
  \put(-2.8,0){${\cal G}\ \ \ =$}
  \put(0,0){$1$}
  \put(2.4,0){$1$}
  \put(0.8,0.5){$\vvdots$}
  \put(1.8,0.9){$1$}
  \put(2.7,0.9){$-1$}
  \put(-0.9,-0.9){$-1$}
  \put(0.6,-0.9){$1$}
  \put(1.5,-0.6){$\vvdots$}
  \end{picture}
  }
  \label{gl}
  \een
  since it adds only {\em one} to the internal weight and not two. This 
  discrepancy follows from the fact that the Dynkin labels satisfy
  $\la_1+\mu_1+\nu_1\in2\Z_\geq$, so if two weights are changed simultaneously
  and equally, we can only require an even change of the {\em sum} of the
Dynkin labels.

  In general, the number of independent loop-gluing diagrams 
is $r$ -- one for each of the Dynkin labels 
that must be identified along the self-gluing. To keep 
  the presentation simple, we list here the three diagrams associated 
  to $su(4)$:
  \ben
  \mbox{
  \begin{picture}(200,115)(80,-55)
  \unitlength=0.6cm
  \thicklines
  \put(-2,0){${\cal L}_1\ \ $=}
  \put(0,2.5){0}
  \put(0,1.5){0}
  \put(0,0.5){0}
  \put(0,-0.5){0}
  \put(0,-1.5){0}
  \put(0,-2.5){0}
  \put(1,2){0}
  \put(1,0){0}
  \put(1,-2){0}
  \put(2,1.5){0}
  \put(2,0.5){0}
  \put(2,-0.5){0}
  \put(2,-1.5){0}
  \put(3,1){0}
  \put(3,-1){0}
  \put(4,0.5){0}
  \put(4,-0.5){0}
  \put(5,0){1}
  \put(7,0){${\cal L}_2\ \ $=}
  \put(9,2.5){0}
  \put(9,1.5){0}
  \put(9,0.5){0}
  \put(9,-0.5){0}
  \put(9,-1.5){0}
  \put(9,-2.5){0}
  \put(10,2){0}
  \put(10,0){0}
  \put(10,-2){0}
  \put(11,1.5){0}
  \put(11,0.5){0}
  \put(11,-0.5){0}
  \put(11,-1.5){0}
  \put(12,1){1}
  \put(12,-1){1}
  \put(13,0.5){0}
  \put(13,-0.5){0}
  \put(14,0){0}
  \put(16,0){${\cal L}_3\ \ $=}
  \put(18,2.5){0}
  \put(18,1.5){0}
  \put(18,0.5){0}
  \put(18,-0.5){0}
  \put(18,-1.5){0}
  \put(18,-2.5){0}
  \put(19,2){1}
  \put(19,0){1}
  \put(19,-2){1}
  \put(20,1.5){0}
  \put(20,0.5){0}
  \put(20,-0.5){0}
  \put(20,-1.5){0}
  \put(21,1){0}
  \put(21,-1){0}
  \put(22,0.5){0}
  \put(22,-0.5){0}
  \put(23,0){0}
  \end{picture}
  }
  \label{gl4}
  \een
The extension to higher rank is obvious. It amounts to introducing a diagram 
with ones in a vertical line (when the triangle is tilted as in (\ref{gl4})), 
while all other entries are zero. The $r$ relevant vertical lines are the 
first, the third, the fifth, etc, when counting from the rightmost vertex.

We are now in a position to discuss general genus-$h$ ${\cal N}$-point 
fusions. The only missing information is how to assign explicitly a threshold 
level to a generic $su(N)$ BZ triangle. As already mentioned, that is known 
for $N\leq4$, so in the following we will focus on $su(3)$ and $su(4)$. We 
will also allude to the straightforward but technically elaborate extension to
higher rank. For results on $su(2)$, we refer to \cite{RW3}.

  \section{Higher-genus $su(3)$ fusion}

  A generic $su(3)$ BZ triangle may be written
  \ben
   \matrix{\quad\cr m_{13}\cr
          n_{12}~~\quad l_{23}\cr
   m_{23}~\quad\qquad ~~m_{12}\cr
   n_{13}~\quad l_{12} \qquad n_{23} \quad~ l_{13} \cr \quad\cr}
  \label{3}
  \een
  with outer constraints
  \bea
   m_{13}+n_{12}=\lambda_1\ \ \ \ \ n_{13}+l_{12}=\mu_1\ \ \ \ \
   l_{13}+m_{12}=\nu_1\nn
   m_{23}+n_{13}=\lambda_2\ \ \ \ \ n_{23}+l_{13}=\mu_2\ \ \ \ \
   l_{23}+m_{13}=\nu_2
  \label{3o}
  \eea
  The hexagon identities are
  \bea
   n_{12}+m_{23}&=&n_{23}+m_{12}\nn 
   m_{12}+l_{23}&=&m_{23}+l_{12}\nn 
   l_{12}+n_{23}&=&l_{23}+n_{12}
  \label{3h}
  \eea
  of which only two are independent.
  The threshold level assigned to (\ref{3}) is \cite{KMSW}
  \ben
   t=\max\{\la_1+\la_2+l_{13},\ \mu_1+\mu_2+m_{13},\ \nu_1+\nu_2+n_{13}\}
  \label{3t}
  \een
  Denoting the level of the affine $su(3)$ by $k$, the affine condition
  \ben
   t\leq k
  \label{affine}
  \een
  supplements the inequalities defining the tensor product polytope. Thus, 
  the discretized volume of the convex polytope
  defined by the inequalities in $v$
\bea
 0&\leq&\la^1-\la^2-\mu^1+\nu^2+v,
  \ \la^1+\mu^1-\nu^2-v,\ \la_2-v,\ v,\ \mu_1-v,\nn
 && \la^2+\mu^2-\nu^1-v,\ -\la^2-\mu^1+\mu^2+\nu^1+v,\ 
  \la^2+\mu^1-\mu^2-\nu^1+\nu^2-v,\nn 
 &&-\la^1+\la^2+\mu^1-\nu^1+\nu^2-v,\nn 
 && k-\la^1+\mu^1-\mu^2-\nu^1-v,\ 
   k-\la^1+\la^2-\mu^2-\nu^2-v,\ k-\nu_1-\nu_2-v
  \label{cp3}
  \eea
  is the fusion multiplicity $N_{\la,\mu,\nu}^{(k)}$. The choice of initial
  diagram ${\cal D}_0$ is implicitly given in (\ref{cp3}), and 
  the volume is easily measured explicitly \cite{RW4}.  
The dual Dynkin labels $\la^i,\ \mu^i$ and $\nu^i$ can be written 
in terms of ordinary Dynkin labels as $\la^1=\frac{1}{3}(2\la_1+\la_2)$
and $\la^2=\frac{1}{3}(\la_1+2\la_2)$, and similarly for $\mu^i$ and $\nu^i$. 
The weights are subject to the condition 
  \bea
  \la^i+\mu^i+\nu^i\in\Z_\geq,\ \ \ i=1,2
  \eea
    
  Let us now consider genus-1 ${\cal N}$-point $su(3)$ fusion multiplicities
  $N_{\la^{(1)},...,\la^{({\cal N})}}^{(k,1)}$. There is a threshold level
  associated to each participating triangle. Thus, there is a 
  level-dependent inequality like (\ref{affine}) associated to each triangle.
  For $h={\cal N}=1$, the tadpole (\ref{tad}) with outer weight $\la$
  may be expressed as
  \ben
   {\cal D}={\cal D}_0+v{\cal V}-\sum_{i=1}^2l_i{\cal L}_i
  \een
  The associated convex polytope (in $v$, $l_1$ and $l_2$) is defined by
  \bea
   0&\leq&\frac{1}{3}(2\la_1+\la_2)+v,\  \frac{1}{3}(\la_1-\la_2)-v,\  
    -\frac{1}{3}(\la_1-\la_2)-v,\ \frac{1}{3}(\la_1+2\la_2)+v,\nn
   &&\frac{1}{3}(\la_1-\la_2)-v+l_2,\ -v,\ v+l_1,\ -v+l_2,\nn
   &&k-\la_1-\la_2-l_1-l_2-v,\ k-\la_1-\la_2-v-l_1,\ 
     k-\frac{1}{3}(4\la_1+2\la_2)-l_1-l_2-v
   \label{1p}
  \eea
  The explicit choice of initial diagram ${\cal D}_0$ is easily read off
  (\ref{1p}). It follows that
  the multiplicity $N_\la^{(k,1)}$ can be written in simplified form as
  \bea
   N_\la^{(k,1)}=\hf(\min\{\la_1,\la_2\}+1)(k+2-\max\{\la_1,\la_2\})
   (k+1-\la_1-\la_2)
  \label{sum1}
  \eea
with $\la^1\in\Z_\geq$ and $\la_1+\la_2\leq k$. The multiplicity
$N_\la^{(k,1)}$ vanishes if these conditions are not satisfied.
Note that $\la^1\in\Z_\geq$ implies $\la^2\in\Z_\geq$.

We can use (\ref{sum1}) to calculate the multiplicity of the genus-1
${\cal N}$-point fusion. It is convenient to  
distinguish between even ${\cal N}$ 
and odd ${\cal N}$. The reason for this is that the triangle corresponding to 
the last point or outer weight $\la^{({\cal N})}$
can have two different orientations before it is glued to the 
tadpole. The following diagram shows the genus-1 ${\cal N}$-point fusion
(in this example ${\cal N}$ is assumed even):  
  \vspace{0.4cm}
  \ben
  \mbox
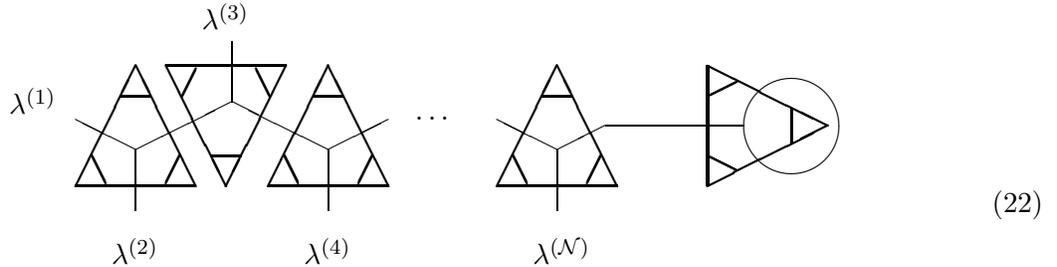
{
  \begin{picture}(180,80)(50,-10)
  \unitlength=0.8cm
  \thicklines
   \put(0,0){\line(1,2){1}}
   \put(0,0){\line(1,0){2}}
   \put(2,0){\line(-1,2){1}}
   \put(0.5,0.05){\line(-1,2){0.23}}
   \put(1.5,0.05){\line(1,2){0.23}}
   \put(0.75,1.5){\line(2,0){0.5}}
  \thinlines
  \put(1,0.6){\line(0,-1){1}}
  \put(1,0.6){\line(2,1){1.6}}
  \put(1,0.6){\line(-2,1){1}}
  \thicklines
  \put(-1.1,1.2){$\la^{(1)}$}
  \put(0.6,-1.3){$\la^{(2)}$}
   \put(1.34,2){
  \begin{picture}(50,50)
   \put(0,0){\line(1,-2){1}}
   \put(0,0){\line(1,0){2}}
   \put(2,0){\line(-1,-2){1}}
   \put(0.5,-0.05){\line(-1,-2){0.23}}
   \put(1.5,-0.05){\line(1,-2){0.23}}
   \put(0.75,-1.5){\line(2,0){0.5}}
  \thinlines
  \put(1.1,-0.6){\line(0,1){1}}
  \thicklines
  \put(0.6,0.6){$\la^{(3)}$}
  \end{picture}}
  \put(3.2,0){\begin{picture}(50,50)
   \put(0,0){\line(1,2){1}}
   \put(0,0){\line(1,0){2}}
   \put(2,0){\line(-1,2){1}}
   \put(0.5,0.05){\line(-1,2){0.23}}
   \put(1.5,0.05){\line(1,2){0.23}}
   \put(0.75,1.5){\line(2,0){0.5}}
  \thinlines
  \put(1,0.6){\line(0,-1){1}}
  \put(1,0.6){\line(2,1){1}}
  \put(1,0.6){\line(-2,1){1.6}}
  \thicklines
  \put(2.45,1.1){$\dots$}
  \put(0.6,-1.3){$\la^{(4)}$}
  \end{picture}}
  \put(7,0){\begin{picture}(50,50)
   \put(0,0){\line(1,2){1}}
   \put(0,0){\line(1,0){2}}
   \put(2,0){\line(-1,2){1}}
   \put(0.5,0.05){\line(-1,2){0.23}}
   \put(1.5,0.05){\line(1,2){0.23}}
   \put(0.75,1.5){\line(2,0){0.5}}
  \thinlines
  \put(1,0.6){\line(0,-1){1}}
  \put(1,0.6){\line(2,1){0.8}}
  \put(1,0.6){\line(-2,1){1}}
  \thicklines
  \put(0.6,-1.3){$\la^{({\cal N})}$}
  \end{picture}}
  \put(10.5,1){\begin{picture}(100,80)
  \unitlength=1.6cm
  \put(0,0.5){\line(0,-1){1}}
  \put(0,0.5){\line(2,-1){1}}
  \put(0,-0.5){\line(2,1){1}}
  \put(0.01,0.25){\line(2,1){0.24}}
  \put(0.01,-0.25){\line(2,-1){0.24}}
  \put(0.7,-0.14){\line(0,2){0.29}}
  \thinlines
  \put(0.7,0){\circle{0.8}}
  \put(0.3,0){\line(-1,0){1.15}}
  \thicklines
  \end{picture}}
  \end{picture}
  }
  \label{even}
  \vspace{0.9cm}
  \een
  Let us introduce the parameter ${\cal B}$ counting -- from the left of 
the diagram -- the number of
pairs of triangles not involving $\la^{(1)}$, $\la^{(2)}$, $\la^{(3)}$ or
$\la^{({\cal N})}$: ${\cal N}=4+2{\cal B}$, i.e.   
  ${\cal B}=\hf{({\cal N}-4)}$. We also introduce the abbreviation
  \ben
   \la^{(1,m)}=\la^{(1)}+\la^{(2)}+...+\la^{(m)}
  \een
  The diagram associated to (\ref{even}) may be written
  \ben
   {\cal D}={\cal D}_0+\sum_{i=0}^{{\cal B}+1}v_i{\cal V}_i+
  \sum_{j=1}^{{\cal B}+1}u_j{\cal U}_j-\sum_{k=1}^{{\cal B}+2}s_k{\cal S}_k-
  \sum_{l=1}^{{\cal B}+2}r_l{\cal R}_l-\sum_{m=1}^{{\cal B}+1}g_m{\cal G}_m-
  \sum_{n=1}^{{\cal B}+1}f_n{\cal F}_n
  \een
A ${\cal V}$ represents a virtual triangle associated to an
upward-pointing triangle, with ${\cal V}_0$ being associated
to the leftmost triangle. Also labeling from the left, a ${\cal U}$
represents a virtual triangle associated to a downward-pointing triangle.
${\cal R}$ and ${\cal S}$ are the gluing diagrams to the right of an
upward-pointing triangle, ${\cal R}$ being the upper one, while
${\cal F}$ and ${\cal G}$ are the gluing diagrams to the right of
a downward-pointing triangle, ${\cal F}$ being the upper one.
The labeling is again from left to right.
  This results in the following polytope-defining list of inequalities:
  \bea
   0&\leq&\la_1^{(2)}+v_0-s_1,\ s_1-v_0,\ \la_2^{(2)}-v_0,\
     k-\la_1^{(2)}-\la_2^{(2)}-\la_2^{(1)}+r_1-v_0\nn
   0&\leq&v_0,\ \la_1^{(1)}-v_0,\ r_1-v_0,\
    k-\la_1^{(1)}-\la_2^{(1)}-\la_1^{(2)}+s_1-v_0\nn
   0&\leq&\la_2^{(1)}+v_0-r_1,\ \la_2^{(2)}-v_0-r_1+s_1,\ \la_1^{(1)}-v_0+r_1
    -s_1,\nn
    &&k-\la_2^{(1)}-\la_2^{(2)}-\la_1^{(2)}-\la_1^{(1)}+r_1+s_1-v_0\nn 
   &\vdots&\nn
   0&\leq&\la_1^{(2b+2)}+v_b-s_{b+1},\ s_{b+1}-v_b,\ \la_2^{(2b+2)}-v_b+g_b,
\nn
    &&k-\la_2^{(1,2b+2)}-\la_1^{(1,2b+2)}+r_{b+1}+s_{b+1}-v_b+g_b\nn
   0&\leq&v_b-g_b,\ \la_1^{(1,2b+1)}-v_b-g_b+f_b,\ g_b-v_b-f_b+r_{b+1},\ \nn
    &&k-\la_1^{(2b+2)}-\la_2^{(1,2b+2)}-v_b+f_b+r_{b+1}\nn
   0&\leq&\la_2^{(1,2b+1)}+v_b-f_b-r_{b+1},\ \la_2^{(2b+2)}-v_b+f_b-r_{b+1}+
    s_{b+1},\ \la_1^{(1,2b+1)}-v_b+r_{b+1}-s_{b+1},\nn
    &&k-\la_1^{(1,2b+2)}-\la_2^{(1,2b+1)}+g_b+f_b-v_b+s_{b+1}\nn
   0&\leq&\la_2^{(2b+1)}+u_b-f_b,\ \la_2^{(1,2b)}-u_b-f_b+g_b,
\ \la_1^{(2b+1)}-
    u_b+f_b-g_b+s_b,\nn 
    &&k-\la_1^{(1,2b+1)}-\la_2^{(1,2b+1)}+f_b+g_b-u_b+r_b\nn
   0&\leq&\la_1^{(1,2b)}+u_b-g_b-s_b,\ g_b-u_b-s_b+r_b,
\ \la_2^{(1,2b)}-u_b-r_b
    +s_b,\nn
    &&k-\la_1^{(1,2b)}-\la_2^{(1,2b+1)}+s_b+r_b-u_b+f_b\nn
   0&\leq&u_b-r_b,\ \la_1^{(2b+1)}-u_b+r_b,\ f_b-u_b, \
    k-\la_1^{(1,2b+1)}-\la_2^{(2b+1)}-u_b+g_b+s_b\nn
   &\vdots&\nn
   0&\leq&\la_1^{({\cal N})}+v_{{\cal B}+1}-s_{{\cal B}+2},\ s_{{\cal B}+2}
-v_{{\cal B}+1},\ \la_2^{({\cal N})}-v_{{\cal B}+1}+g_{{\cal B}+1}, \nn
   &&k-\la_1^{({\cal N})}-\la_2^{(1,{\cal N})}-v_{{\cal B}+1}
  +f_{{\cal B}+1}+r_{{\cal B}+2}\nn
 0&\leq&v_{{\cal B}+1}-g_{{\cal B}+1},\ \la_1^{(1,{\cal N}-1)}-v_{{\cal B}+1}-
   g_{{\cal B}+1}+f_{{\cal B}+1},\ g_{{\cal B}+1}-v_{{\cal B}+1}-f_
    {{\cal B}+1}+r_{{\cal B}+2}, \nn
 &&k-\la_1^{(1,{\cal N})}-\la_2^{(1,{\cal N}-1)}-v_{{\cal B}+1}+f_{{\cal B}+1}
   +g_{{\cal B}+1}+s_{{\cal B}+2} \nn
 0&\leq&\la_2^{(1,{\cal N}-1)}+v_{{\cal B}+1}-f_{{\cal B}+1}-r_{{\cal B}+2},\ 
   \la_2^{({\cal N})}-v_{{\cal B}+1}-r_{{\cal B}+2}+s_{{\cal B}+2}+f_{{\cal B}
   +1},\nn
 & & \la_1^{(1,{\cal N}-1)}-v_{{\cal B}+1}+r_{{\cal B}+2}-s_{{\cal B}+2},\
    k-\la_1^{(1,{\cal N})}-\la_2^{(1,{\cal N})}-v_{{\cal B}+1}
     +g_{{\cal B}+1}+s_{{\cal B}+2}+r_{{\cal B}+2}\nn
   0&\leq&\la_2^{({\cal N}-1)}+u_{{\cal B}+1}-f_{{\cal B}+1},\ \la_2^{(1,
     {\cal N}-2)}-u_{{\cal B}+1}-f_{{\cal B}+1}+g_{{\cal B}+1},\nn
  && \la_1^{({\cal N}-1)}-u_{{\cal B}+1}-g_{{\cal B}+1}+f_{{\cal B}+1}+
    s_{{\cal B}+1},\nn
& &k-\la_2^{(1,{\cal N}-1)}-\la_1^{(1,{\cal N}-1)}-u_{{\cal B}+1}+f_{{\cal B}+
  1}+g_{{\cal B}+1}+r_{{\cal B}+1}\nn
 0&\leq&\la_1^{(1,{\cal N}-2)}+u_{{\cal B}+1}-g_{{\cal B}+1}-s_{{\cal B}+1},\ 
    g_{{\cal B}+1}-u_{{\cal B}+1}-s_{{\cal B}+1}+r_{{\cal B}+1},\nn
  && \la_2^{(1,{\cal N}-2)}-u_{{\cal B}+1}+s_{{\cal B}+1}-r_{{\cal B}+1},\
    k-\la_1^{(1,{\cal N}-2)}-\la_2^{(1,{\cal N}-1)}+f_{{\cal B}+1}+s_{{\cal B}
    +1}+r_{{\cal B}+1}-u_{{\cal B}+1}\nn
   0&\leq&u_{{\cal B}+1}-r_{{\cal B}+1},\ \la_1^{({\cal N}-1)}-u_{{\cal B}+1}+
  r_{{\cal B}+1},\ f_{{\cal B}+1}-u_{{\cal B}+1},\nn
& &k-\la_1^{(1,{\cal N}-1)}-\la_2^{({\cal N}-1)}+g_{{\cal B}+1}+s_{{\cal B}+1}
  -u_{{\cal B}+1}
  \label{polgN}
  \eea
  Here $b$ is a label defined in the interval $1\leq b\leq{\cal B}$.
  The volume may be measured explicitly expressing 
  $N_{\la^{(1)},...,\la^{({\cal N})}}^{(k,1)}$ as a multiple sum, with the 
  appropriate order of summation:
  \bea
  N_{\la^{(1)},...,\la^{({\cal N})}}^{(k,1)}&=& 
   \left(\sum_{v_0}\sum_
    {s_1}\sum_{r_1}\right)\sum_{b}\left(\sum_{u_b}\sum_
    {g_b}\sum_{f_b}\sum_{v_b}\sum_{s_b+1}\sum_{r_b+1}\right)\nn
  &\times&\sum_{u_{{\cal B}+
    1}}\sum_{g_{{\cal B}+1}}\sum_{f_{{\cal B}+1}}\sum_{v_{{\cal B}+1}}\sum_{
    s_{{\cal B}+2}}\sum_{r_{{\cal B}+2}}
  N_{\la^{(1,{\cal N})}-s_{{\cal B}}\al_1-r_{{\cal B}}\al_2}^{(k,1)}
  \label{sumNh}
  \eea
  According to (\ref{sum1}), the summand
  $N_{\la^{(1,{\cal N})}-s_{{\cal B}}\al_1-r_{{\cal B}}\al_2}^{(k,1)}$ 
may be expressed as
  \bea
  N_{\la^{(1,{\cal N})}-s_{{\cal B}}\al_1-r_{{\cal B}}\al_2}^{(k,1)}
   &=&\hf\left(\min\{\la_1^{(1,{\cal N})}-
   2s_{{\cal B}+2}+r_{{\cal B}+2},\la_2^
    {(1,{\cal N})}+s_{{\cal B}+2}-2r_{{\cal B}+2}\}+1\right)\nn
   &\times&\left(k+2-\max\{\la_1^{(1,{\cal N})}-2s_{{\cal B}+2}
    +r_{{\cal B}+2},\la_2^{(1,{\cal N})}+s_{{\cal B}+2}
    -2r_{{\cal B}+2}\}\right)\nn
   &\times&\left(k+1-\la_1^{(1,{\cal N})}
    -\la_2^{(1,{\cal N})}+s_{{\cal B}+2}+r_{{\cal B}+2}\right)
  \label{sum2}
  \eea
  The lower bounds on the summation variables are 
  \bea
   v_0&\geq&0\nn
   s_1&\geq&\max\{\la_1^{(1)}+\la_2^{(1)}+\la_1^{(2)}+v_0-k,\ v_0\}\nn
   r_1&\geq&\max\{v_0,\ v_0-\la_1^{(1)}+s_1,\ \la_1^{(2)}+\la_2^{(2)}+\la_2^
     {(1)}+v_0-k,\nn
  && \la_2^{(1)}+\la_1^{(1)}+\la_2^{(2)}+\la_1^{(2)}-s_1+v_0-k\},\nn
   b&\geq&1\nn
   u_b&\geq&r_b\nn
   g_b&\geq&\max\{u_b+s_b-r_b,\ \la_1^{(1,2b+1)}+\la_2^{(2b+1)}+u_b-s_b-k\}\nn
   f_b&\geq&\max\{\la_1^{(1,2b)}+\la_2^{(1,2b+1)}-s_b-r_b+u_b-k,\ \la_1^
     {(1,2b+1)}+\la_2^{(1,2b+1)}-g_b+u_b-r_b-k,\nn
  && u_b,\ u_b-\la_1^{(2b+1)}+g_b-s_b\}\nn
   v_b&\geq&g_b\nn
   s_{b+1}&\geq&\max\{v_b,\ \la_1^{(1,2b+2)}+\la_2^{(1,2b+1)}-g_b-f_b+v_b-k\}
    \nn
   r_{b+1}&\geq&\max\{\la_1^{(2b+2)}+\la_2^{(1,2b+2)}+v_b-f_b-k,\ \la_1^
     {(1,2b+2)}+\la_2^{(1,2b+2)}-s_{b+1}+v_b-g_b-k,\nn
   & & v_b-\la_1^{(1,2b+1)}+s_{b+1},\ v_b+f_b-g_b\}\nn
  u_{{\cal B}+1}&\geq&r_{{\cal B}+1}\nn 
   g_{{\cal B}+1}&\geq&\max\{\la_1^{(1,{\cal N}-1)}+\la_2^{({\cal N}-1)}+u_
     {{\cal B}+1}-s_{{\cal B}+1}-k,\ u_{{\cal B}+1}+s_{{\cal B}+1}-r_
     {{\cal B}+1}\}\nn
  f_{{\cal B}+1}&\geq&\max\{\la_1^{(1,{\cal N}-2)}+\la_2^{(1,{\cal N}-1)}-s_
     {{\cal B}+1}-r_{{\cal B}+1}+u_{{\cal B}+1}-k,\ \la_1^{(1,{\cal N}-1)}+
     \la_2^{(1,{\cal N}-1)}\nn
  &&-g_{{\cal B}+1}+u_{{\cal B}+1}-r_{{\cal B}+1}-k,\ u_{{\cal B}+1},\
   u_{{\cal B}+1}-\la_1^{({\cal N}-1)}+g_{{\cal B}+1}-s_{{\cal B}+1}\}\nn
  v_{{\cal B}+1}&\geq&g_{{\cal B}+1}\nn
  s_{{\cal B}+2}&\geq&\max\{\la_1^{(1,{\cal N})}+\la_2^{(1,{\cal N}-1)}+v_
     {{\cal B}+1}-f_{{\cal B}+1}-g_{{\cal B}+1}-k,\ v_{{\cal B}+1}\}\nn
  r_{{\cal B}+2}&\geq&\max\{\la_1^{(1,{\cal N})}+\la_2^{(1,{\cal N})}+v_
     {{\cal B}+1}-g_{{\cal B}+1}-s_{{\cal B}+2}-k,\ 
     v_{{\cal B}+1}-\la_1^{(1,{\cal N}-1)}+s_{{\cal B}+2},\nn
  && \la_1^{({\cal N})}+\la_2^{(1,{\cal N})}+v_{{\cal B}+1}-f_{{\cal B}+1}-k,
     \ v_{{\cal B}+1}-g_{{\cal B}+1}+f_{{\cal B}+1}\}
  \label{sum3}
  \eea
  The upper bounds of the summation variables are
  \bea
   v_0&\leq&\min\{\la_1^{(1)},\ \la_2^{(2)}\}\nn
   s_1&\leq&\la_1^{(2)}+v_0\nn
   r_1&\leq&\min\{\la_2^{(1)}+v_0,\ \la_2^{(2)}-v_0+s_1\}\nn
   b&\leq&{\cal B}\nn
   u_b&\leq&\min\{\la_1^{(2b+1)}+r_b,\ \la_2^{(1,2b)}-r_b+s_b\}\nn
   g_b&\leq&\la_1^{(1,2b)}+u_b-s_b\nn
   f_b&\leq&\min\{\la_2^{(1,2b)}-u_b+g_b,\ \la_2^{(2b+1)}+u_b\}\nn
   v_b&\leq&\min\{\la_1^{(1,2b+1)}-g_b+f_b,\ \la_2^{(2b+2)}+g_b\}\nn
   s_{b+1}&\leq&\la_1^{(2b+2)}+v_b\nn
   r_{b+1}&\leq&\min\{\la_2^{(1,2b+1)}+v_b-f_b,\ \la_2^{(2b+2)}-v_b
  +f_b+s_{b+1}\}\nn
   u_{{\cal B}+1}&\leq&\min\{\la_2^{(1,{\cal N}-2)}+s_{{\cal B}+1}
   -r_{{\cal B}+
     1},\ \la_1^{({\cal N}-1)}+r_{{\cal B}+1}\}\nn 
 g_{{\cal B}+1}&\leq&\la_1^{(1,{\cal N}-2)}+ u_{{\cal B}+1}- s_{{\cal B}+1}\nn
  f_{{\cal B}+1}&\leq&\min\{\la_2^{(1,{\cal N}-2)}- u_{{\cal B}+1}
   + g_{{\cal B}
     +1},\ \la_2^{({\cal N}-1)}+ u_{{\cal B}+1}\}\nn
v_{{\cal B}+1}&\leq&\min\{\la_2^{({\cal N})}+g_{{\cal B}+1},\ \la_1^{(1,{\cal 
    N}-1)}-g_{{\cal B}+1}+f_{{\cal B}+1}\}\nn
  s_{{\cal B}+2}&\leq&\la_1^{({\cal N})}+v_{{\cal B}+1}\nn
r_{{\cal B}+2}&\leq&\min\{\la_2^{(1,{\cal N}-1)}+v_{{\cal B}+1}-f_{{\cal B}+1}
    ,\ \la_2^{({\cal N})}-v_{{\cal B}+1}+s_{{\cal B}+2}+f_{{\cal B}+1}\}
  \label{sum4}
  \eea

  The fusion multiplicity for ${\cal N}$ odd may be computed similarly.
The associated diagram 
  is similar to the even case (\ref{even}) except that the second triangle
  from the right is turned upside down:
  \ben
  \mbox
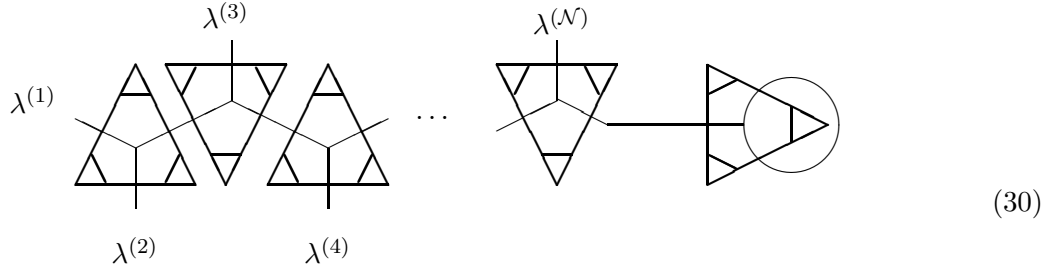
{
  \begin{picture}(180,80)(50,-10)
  \unitlength=0.8cm
  \thicklines
   \put(0,0){\line(1,2){1}}
   \put(0,0){\line(1,0){2}}
   \put(2,0){\line(-1,2){1}}
   \put(0.5,0.05){\line(-1,2){0.23}}
   \put(1.5,0.05){\line(1,2){0.23}}
   \put(0.75,1.5){\line(2,0){0.5}}
  \thinlines
  \put(1,0.6){\line(0,-1){1}}
  \put(1,0.6){\line(2,1){1.6}}
  \put(1,0.6){\line(-2,1){1}}
  \thicklines
  \put(-1.1,1.2){$\la^{(1)}$}
  \put(0.6,-1.3){$\la^{(2)}$}
   \put(1.34,2){
  \begin{picture}(50,50)
   \put(0,0){\line(1,-2){1}}
   \put(0,0){\line(1,0){2}}
   \put(2,0){\line(-1,-2){1}}
   \put(0.5,-0.05){\line(-1,-2){0.23}}
   \put(1.5,-0.05){\line(1,-2){0.23}}
   \put(0.75,-1.5){\line(2,0){0.5}}
  \thinlines
  \put(1.1,-0.6){\line(0,1){1}}
  \thicklines
  \put(0.6,0.6){$\la^{(3)}$}
  \end{picture}}
  \put(3.2,0){\begin{picture}(50,50)
   \put(0,0){\line(1,2){1}}
   \put(0,0){\line(1,0){2}}
   \put(2,0){\line(-1,2){1}}
   \put(0.5,0.05){\line(-1,2){0.23}}
   \put(1.5,0.05){\line(1,2){0.23}}
   \put(0.75,1.5){\line(2,0){0.5}}
  \thinlines
  \put(1,0.6){\line(0,-1){1}}
  \put(1,0.6){\line(2,1){1}}
  \put(1,0.6){\line(-2,1){1.6}}
  \thicklines
  \put(2.45,1.1){$\dots$}
  \put(0.6,-1.3){$\la^{(4)}$}
  \end{picture}}
  \put(7,0){\begin{picture}(50,50)
   \put(1,0){\line(1,2){1}}
   \put(0,2){\line(1,0){2}}
   \put(1,0){\line(-1,2){1}}
   \put(1.7,1.5){\line(-1,2){0.23}}
   \put(0.3,1.5){\line(1,2){0.23}}
   \put(0.75,0.5){\line(2,0){0.5}}
  \thinlines
  \put(1,2.4){\line(0,-1){1}}
  \put(0,0.9){\line(2,1){1}}
  \put(1.85,1){\line(-2,1){0.85}}
  \thicklines
  \put(0.6,2.5){$\la^{({\cal N})}$}
  \end{picture}}
  \put(10.5,1){\begin{picture}(100,80)
  \unitlength=1.6cm
  \put(0,0.5){\line(0,-1){1}}
  \put(0,0.5){\line(2,-1){1}}
  \put(0,-0.5){\line(2,1){1}}
  \put(0.01,0.25){\line(2,1){0.24}}
  \put(0.01,-0.25){\line(2,-1){0.24}}
  \put(0.7,-0.14){\line(0,2){0.29}}
  \thinlines
  \put(0.7,0){\circle{0.8}}
  \put(0.3,0){\line(-1,0){1.13}}
  \thicklines
  \end{picture}}
  \end{picture}
  }
  \label{odd}
  \vspace{0.9cm}
  \een
 In this case, we let ${\cal B}$ count the number of pairs of triangles
not involving $\la^{(1)}$, $\la^{(2)}$ or $\la^{({\cal N})}$:
${\cal N}=3+2{\cal B}$, i.e. ${\cal B}=\hf{({\cal N}-3)}$. 
 Listing the inequalities, we have the following convex polytope:
  \bea
   0&\leq&\la_1^{(2)}+v_0-s_1,\ s_1-v_0,\ \la_2^{(2)}-v_0,\
    k-\la_1^{(2)}-\la_2^{(2)}-\la_2^{(1)}+r_1-v_0 ,\nn
   0&\leq&v_0,\ \la_1^{(1)}-v_0,\ r_1-v_0,\
    k-\la_1^{(1)}-\la_2^{(1)}-\la_1^{(2)}+s_1-v_0,\nn
   0&\leq&\la_2^{(1)}+v_0-r_1,\ \la_2^{(2)}-v_0-r_1+s_1,\ \la_1^{(1)}-v_0+r_1
     -s_1,\nn
    &&k-\la_2^{(1)}-\la_2^{(2)}-\la_1^{(2)}-\la_1^{(1)}+r_1+s_1-v_0\nn 
   &\vdots&\nn
   0&\leq&\la_1^{(2b+2)}+v_b-s_{b+1},\ s_{b+1}-v_b,\ \la_2^{(2b+2)}
-v_b+g_b,\nn
    &&k-\la_2^{(1,2b+2)}-\la_1^{(1,2b+2)}+r_{b+1}+s_{b+1}-v_b+g_b\nn
   0&\leq&v_b-g_b,\ \la_1^{(1,2b+1)}-v_b-g_b+f_b,\ g_b-v_b-f_b+r_{b+1},\ \nn
    &&k-\la_1^{(2b+2)}-\la_2^{(1,2b+2)}-v_b+f_b+r_{b+1}\nn
   0&\leq&\la_2^{(1,2b+1)}+v_b-f_b-r_{b+1},\ \la_2^{(2b+2)}-v_b+f_b-r_{b+1}+
    s_{b+1},\nn
  &&\la_1^{(1,2b+1)}-v_b+r_{b+1}-s_{b+1},\ k-\la_1^{(1,2b+2)}-\la_2^{(1,2b+1)}
    +g_b+f_b-v_b+s_{b+1}\nn
   0&\leq&\la_2^{(2b+1)}+u_b-f_b,\ \la_2^{(1,2b)}-u_b-f_b+g_b,\ \la_1^{(2b+1)}
    -u_b+f_b-g_b+s_b,\nn 
    &&k-\la_1^{(1,2b+1)}-\la_2^{(1,2b+1)}+f_b+g_b-u_b+r_b\nn
   0&\leq&\la_1^{(1,2b)}+u_b-g_b-s_b,\ g_b-u_b-s_b+r_b,\ \la_2^{(1,2b)}-u_b-
    r_b+s_b,\nn
    &&k-\la_1^{(1,2b)}-\la_2^{(1,2b+1)}+s_b+r_b-u_b+f_b\nn
   0&\leq&u_b-r_b,\ \la_1^{(2b+1)}-u_b+r_b,\ f_b-u_b, \nn
   &&k-\la_1^{(1,2b+1)}-\la_2^{(2b+1)}-u_b+g_b+s_b\nn
   &\vdots&\nn
   0&\leq&\la_2^{({\cal N})}+u_{{\cal B}+1}-f_{{\cal B}+1},
  \ \la_2^{(1,{\cal N}
     -1)}-u_{{\cal B}+1}-f_{{\cal B}+1}+g_{{\cal B}+1},\nn
  && \la_1^{({\cal N})}-u_{{\cal B}+1}+f_{{\cal B}+1}-g_{{\cal B}+1}+
    s_{{\cal B}+1}, \nn
 &&k-\la_2^{(1,{\cal N})}-\la_1^{(1,{\cal N})}+g_{{\cal B}+1}+f_{{\cal B}+1}  
      -u_{{\cal B}+1}+r_{{\cal B}+1}\nn
 0&\leq&\la_1^{(1,{\cal N}-1)}+u_{{\cal B}+1}-g_{{\cal B}+1}-s_{{\cal B}+1},\ 
     g_{{\cal B}+1}-u_{{\cal B}+1}-s_{{\cal B}+1}+r_{{\cal B}+1},\nn
  && \la_2^{(1,{\cal N}-1)}-u_{{\cal B}+1}+s_{{\cal B}+1}-r_{{\cal B}+1},
  k-\la_1^{({\cal N}-1)}-\la_2^{(1,{\cal N})}+s_{{\cal B}+1}+r_{{\cal B}+1}-
     u_{{\cal B}+1}+f_{{\cal B}+1}\nn
   0&\leq&u_{{\cal B}+1}-r_{{\cal B}+1},\ \la_1^{({\cal N})}-u_{{\cal B}+1}+
      r_{{\cal B}+1},\ f_{{\cal B}+1}-u_{{\cal B}+1}, \nn
   &&k-\la_1^{(1,{\cal N})}-\la_2^{({\cal N})}-u_{{\cal B}+1}+g_{{\cal B}+1}+
     s_{{\cal B}+1}
  \label{polgN2}
  \eea
  As for ${\cal N}$ even, the volume may be measured explicitly expressing 
  $N_{\la^{(1)},...,\la^{({\cal N})}}^{(k,1)}$ as a multiple sum, with the 
  appropriate order of summation:
  \bea
   N_{\la^{(1)},...,\la^{({\cal N})}}^{(k,1)}&=&\left(\sum_{v_0}\sum_
     {s_1}\sum_{r_1}\right)\sum_{b}\left(\sum_{u_b}\sum_
     {g_b}\sum_{f_b}\sum_{v_b}\sum_{s_b+1}\sum_{r_b+1}\right)\nn
 &\times&\sum_{u_{{\cal B}+
     1}}\sum_{g_{{\cal B}+1}}\sum_{f_{{\cal B}+1}}
   N_{\la^{(1,{\cal N})}-g_{{\cal B}+1}\al_1-f_{{\cal B}+1}\al_2}^{(k,1)}
  \label{sum5}
  \eea
  where the genus-1 1-point fusion multiplicity 
  $N_{\la^{(1,{\cal N})}-g_{{\cal B}+1}\al_1-f_{{\cal B}+1}\al_2}^{(k,1)}$ is 
  \bea
   N_{\la^{(1,{\cal N})}-g_{{\cal B}+1}\al_1-f_{{\cal B}+1}\al_2}^{(k,1)}
 &=&\hf\left(\min\{\la_1^{(1,{\cal N})}-
     2g_{{\cal B}+1}+f_{{\cal B}+1},\la_2^
     {(1,{\cal N})}+g_{{\cal B}+1}-2f_{{\cal B}+1}\}+1\right)\nn
   &\times&\left(k+2-\max\{\la_1^{(1,{\cal N})}-2g_{{\cal B}+1}
    +f_{{\cal B}+1},\la_2^{(1,{\cal N})}+g_{{\cal B}+1}-2f_{{\cal B}+1}\}
    \right)\nn
   &\times&\left(k+1-\la_1^{(1,{\cal N})}
    -\la_2^{(1,{\cal N})}+g_{{\cal B}+1}+f_{{\cal B}+1}\right)
  \label{sum6}
  \eea
  The lower bounds of the summations are
  \bea
   v_0&\geq&0\nn
   s_1&\geq&\max\{\la_1^{(1)}+\la_2^{(1)}+\la_1^{(2)}+v_0-k,\ v_0\}\nn
   r_1&\geq&\max\{v_0,\ v_0-\la_1^{(1)}+s_1,\ \la_1^{(2)}+\la_2^{(2)}+\la_2^
      {(1)}+v_0-k,\nn
  && \la_2^{(1)}+\la_1^{(1)}+\la_2^{(2)}+\la_1^{(2)}-s_1+v_0-k\}\nn
   b&\geq&1\nn
   u_b&\geq&r_b\nn
   g_b&\geq&\max\{u_b+s_b-r_b,\ \la_1^{(1,2b+1)}+\la_2^{(2b+1)}+u_b-s_b-k\}\nn
   f_b&\geq&\max\{\la_1^{(1,2b)}+\la_2^{(1,2b+1)}-s_b-r_b+u_b-k,\nn
  &&\la_1^{(1,2b+1)}+\la_2^{(1,2b+1)}-g_b+u_b-r_b-k,\ u_b,\ 
    u_b-\la_1^{(2b+1)}+g_b-s_b\}\nn
   v_b&\geq&g_b\nn
 s_{b+1}&\geq&\max\{v_b,\ \la_1^{(1,2b+2)}+\la_2^{(1,2b+1)}-g_b-f_b+v_b-k\}\nn
   r_{b+1}&\geq&\max\{\la_1^{(2b+2)}+\la_2^{(1,2b+2)}+v_b-f_b-k,\ \la_1^
      {(1,2b+2)}+\la_2^{(1,2b+2)}\nn
  &&-s_{b+1}+v_b-g_b-k,\ v_b-\la_1^{(1,2b+1)}+s_{b+1},\ v_b+f_b-g_b\}\nn
   u_{{\cal B}+1}&\geq&r_{{\cal B}+1}\nn 
 g_{{\cal B}+1}&\geq&\max\{\la_1^{(1,{\cal N})}+\la_2^{({\cal N})}+u_{{\cal B}
    +1}-s_{{\cal B}+1}-k,\ u_{{\cal B}+1}+s_{{\cal B}+1}-r_{{\cal B}+1}\}\nn
f_{{\cal B}+1}&\geq&\max\{\la_1^{(1,{\cal N}-1)}+\la_2^{(1,{\cal N})}-s_{{\cal
     B}+1}-r_{{\cal B}+1}+u_{{\cal B}+1}-k,\nn
  && \la_1^{(1,{\cal N})}+\la_2^{(1,{\cal N})}-g_{{\cal B}+1}+u_{{\cal B}+1}-
    r_{{\cal B}+1}-k,\nn
   & & u_{{\cal B}+1},\ u_{{\cal B}+1}-\la_1^{({\cal N})}+g_{{\cal B}+1}-
     s_{{\cal B}+1}\}
  \label{lower}
  \eea
  The upper bounds of the summations are
  \bea
   v_0&\leq&\min\{\la_1^{(1)},\ \la_2^{(2)}\}\nn
   s_1&\leq&\la_1^{(2)}+v_0\nn
   r_1&\leq&\min\{\la_2^{(1)}+v_0,\ \la_2^{(2)}-v_0+s_1\}\nn
   b&\leq&{\cal B}\nn
   u_b&\leq&\min\{\la_1^{(2b+1)}+r_b,\ \la_2^{(1,2b)}-r_b+s_b\}\nn
   g_b&\leq&\la_1^{(1,2b)}+u_b-s_b\nn
   f_b&\leq&\min\{\la_2^{(1,2b)}-u_b+g_b,\ \la_2^{(2b+1)}+u_b\}\nn
   v_b&\leq&\min\{\la_1^{(1,2b+1)}-g_b+f_b,\ \la_2^{(2b+2)}+g_b\}\nn
   s_{b+1}&\leq&\la_1^{(2b+2)}+v_b\nn
   r_{b+1}&\leq&\min\{\la_2^{(1,2b+1)}+v_b-f_b,\ \la_2^{(2b+2)}-v_b+f_b+s_
     {b+1}\}\nn
   u_{{\cal B}+1}&\leq&\min\{\la_2^{(1,{\cal N}-1)}+s_{{\cal B}+1}-r_{{\cal B}
      +1},\ \la_1^{({\cal N})}+r_{{\cal B}+1}\}\nn 
   g_{{\cal B}+1}&\leq&\la_1^{(1,{\cal N}-1)}+ u_{{\cal B}+1}
-s_{{\cal B}+1}\nn
   f_{{\cal B}+1}&\leq&\min\{\la_2^{(1,{\cal N}-1)}- u_{{\cal B}+1}+ 
    g_{{\cal B}+1},\ \la_2^{({\cal N})}+ u_{{\cal B}+1}\}
  \label{sumNhb}
  \eea

  These results constitute the first explicit results for 
  higher-genus ${\cal N}$-point $su(3)$ fusion multiplicities.  
For general $h>1$, the polytope characterization of $su(3)$ fusion
multiplicities is straightforward, but very cumbersome and will not
be discussed explicitly here.
  In the next section we will consider briefly the extension to
  $su(N)$, while Appendix A contains details on $su(4)$.

We conclude this section by writing down a double-sum formula
for the genus-2 0-point $su(3)$ fusion multiplicity. It is obtained by 
gluing two genus-1 1-point fusions (\ref{sum1}) together, and summing over
the internal weight subject to the conditions following
(\ref{sum1}). We find
\ben
 N^{(k,2)}\ =\ \sum_{i=0}^{[k/2]}\ \sum_{j=0}^{[(k-2i)/3]}\
  \left(\frac{1}{2}(i+1)(k+2-i-3j)(k+1-2i-3j)\right)^2
\een
  where $[x]$ denotes the integer value of $x$, i.e. the greatest integer 
  less than or equal to $x$. This double sum may be summed explicitly,
and we find that the genus-2 0-point $su(3)$ fusion multiplicity
is a simple binomial coefficient in the affine level $k$:
\ben
 N^{(k,2)}\ =\ \bin{k+8}{k}
\een
This is believed to be the first concise result on fusion multiplicities
for $r,h>1$.

  \section{On higher-genus $su(N)$ fusion}

  Pending on the explicit assignment of threshold levels to true BZ triangles
  for $N>4$, our method allows us to characterize all higher-genus 
  ${\cal N}$-point $su(N)$ fusion multiplicities by convex polytopes.
  Let us indicate how by considering the genus-$h$ ${\cal N}$-point
  diagram
  \ben
  \mbox
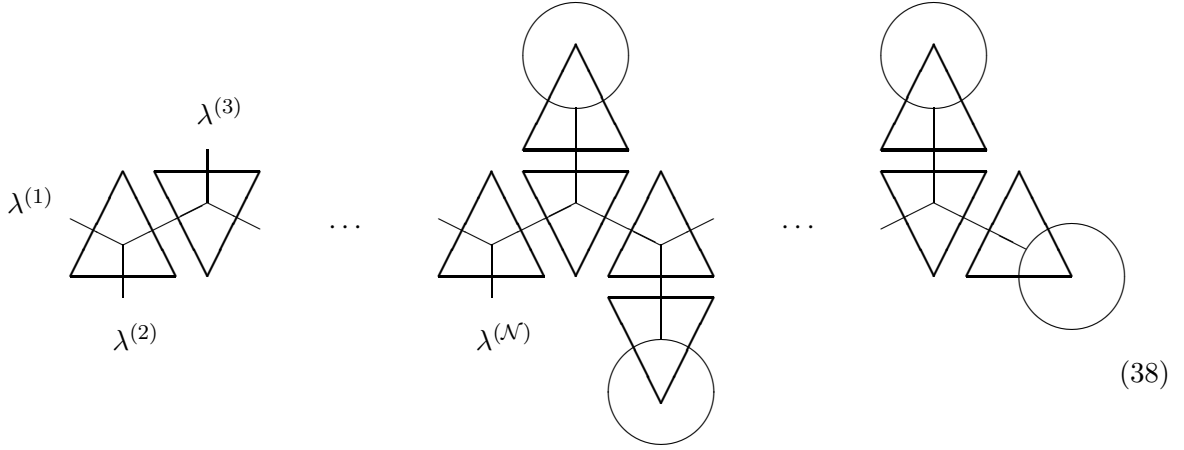
{
  \begin{picture}(180,160)(100,-40)
  \unitlength=0.7cm
  \thicklines
   \put(0,0){\line(1,2){1}}
   \put(0,0){\line(1,0){2}}
   \put(2,0){\line(-1,2){1}}
   \put(1.6,2){\line(1,0){2}}
   \put(1.6,2){\line(1,-2){1}}
   \put(3.6,2){\line(-1,-2){1}}
  \thinlines
   \put(1,0.6){\line(0,-1){1}}
   \put(1,0.6){\line(2,1){1.6}}
   \put(1,0.6){\line(-2,1){1}}
   \put(2.6,1.4){\line(0,1){1}}
   \put(2.6,1.4){\line(2,-1){1}}
   \put(4.9,0.8){$\cdots$}
  \thicklines
   \put(7,0){\line(1,2){1}}
   \put(7,0){\line(1,0){2}}
   \put(9,0){\line(-1,2){1}}
   \put(8.6,2){\line(1,0){2}}
   \put(8.6,2){\line(1,-2){1}}
   \put(10.6,2){\line(-1,-2){1}}
   \put(10.22,0){\line(1,2){1}}
   \put(10.22,0){\line(1,0){2}}
   \put(12.22,0){\line(-1,2){1}}
   \put(13.5,0.8){$\cdots$}
   \put(15.4,2){\line(1,0){2}}
   \put(15.4,2){\line(1,-2){1}}
   \put(17.4,2){\line(-1,-2){1}}
   \put(17.02,0){\line(1,2){1}}
   \put(17.02,0){\line(1,0){2}}
   \put(19.02,0){\line(-1,2){1}}
   \put(8.6,2.4){\line(1,0){2}}
   \put(8.6,2.4){\line(1,2){1}}
   \put(10.6,2.4){\line(-1,2){1}}
   \put(10.22,-0.4){\line(1,-2){1}}
   \put(10.22,-0.4){\line(1,0){2}}
   \put(12.22,-0.4){\line(-1,-2){1}}
   \put(15.4,2.4){\line(1,0){2}}
   \put(15.4,2.4){\line(1,2){1}}

   \put(17.4,2.4){\line(-1,2){1}}
  \thinlines
   \put(8,0.6){\line(0,-1){1}}
   \put(8,0.6){\line(2,1){1.6}}
   \put(8,0.6){\line(-2,1){1}}
   \put(9.6,1.4){\line(0,1){1.8}}
   \put(9.6,1.4){\line(2,-1){1.6}}
   \put(11.22,0.6){\line(0,-1){1.8}}
   \put(11.22,0.6){\line(2,1){1}}
   \put(16.4,1.4){\line(0,1){1.8}}
   \put(16.4,1.4){\line(2,-1){1.75}}
   \put(16.4,1.4){\line(-2,-1){1}}
   \put(9.6,4.2){\circle{2}}
   \put(11.22,-2.2){\circle{2}}
   \put(16.4,4.2){\circle{2}}
   \put(19.02,0){\circle{2}}
   \put(-1.2,1.2){$\la^{(1)}$}
   \put(0.8,-1.4){$\la^{(2)}$}
   \put(2.4,2.9){$\la^{(3)}$}
   \put(7.7,-1.4){$\la^{({\cal N})}$}
  \end{picture}
  }
  \label{gen}
  \vspace{1cm}
  \een
  In this particular example, ${\cal N}$ and $h$ are assumed even.

  For $h,{\cal N}>0$, the number of glued triangles is ${\cal N}+2h-2$.
  Thus, assuming that the assignment of threshold levels, $t_q$, to the
  individual triangles, ${\cal T}_q$, is known, we have the level-dependent
  inequalities
  \ben
   t_q\leq k,\ \ \ \ q=1,..., {\cal N}+2h-2
  \een
  All other conditions follow from demanding that all entries
  must be {\em non-negative} integers. This leads to
  \ben
   ({\cal N}+2h-2)E_r=\frac{3}{2}({\cal N}+2h-2)r(r+1)
  \een
  level-independent inequalities. The entries and threshold levels are
  given in terms of an initial diagram ${\cal D}_0$, the 
  $({\cal N}+2h-2)H_r$ virtual triangle parameters $v$, the 
  $({\cal N}+2h-3)r$ gluing parameters $g$, and the $hr$ loop-gluing 
parameters
  $l$. In a self-explanatory notation, we have
  \ben
   {\cal D}={\cal D}_0+\sum_{a=1}^{{\cal N}+2h-2}\sum_{i,j=1}^{i+j=r}
    v_{i,j}^{(a)}{\cal V}_{i,j}^{(a)}+\sum_{a=1}^{{\cal N}+2h-3}\sum_{i=1}^r
    g_i^{(a)}{\cal G}_i^{(a)}+\sum_{a=1}^h\sum_{i=1}^rl_i^{(a)}
    {\cal L}_i^{(a)}
  \label{Dgen}
  \een
  The sign convention is immaterial. We see that the polytope
  is embedded in the Euclidean space $\R^{r((h-1)(r+2)+{\cal N}(r+1)/2)}$.

  The initial diagram ${\cal D}_0$ depends only on the ${\cal N}$ outer
  weights $\la^{(1)}$,..., $\la^{({\cal N})}$. A convenient choice
  is characterized by having vanishing entries to the right of the
  $({\cal N}-1)$th triangle in (\ref{gen}) (counting from the left). 
  The entries of the remaining ${\cal N}-1$ (leftmost) triangles follow the
  description of genus-0 $({\cal N}+1)$-point diagrams in \cite{RW2},
  with a vanishing $({\cal N}+1)$th weight being located on the triangle edge
  along which we just imagined the diagram (\ref{gen}) to be cut.

  This concludes the characterization of genus-$h$ ${\cal N}$-point
  $su(N)$ fusion multiplicities by polytopes. The discretized volume
  of the polytope associated to a generic higher-genus
  fusion multiplicity is in general not straightforward to measure,
  and will not be addressed further here. By construction, however,
  it gives the fusion multiplicity 
  $N_{\la^{(1)},...,\la^{({\cal N})}}^{(k,h)}$.

  Of particular interest are the genus-1 0-point fusion multiplicities
  $N^{(k,1)}$, which depend solely on the affine level $k$ and the
  rank $r$ of $su(r+1)$. They also
provide a nice check of our general picture. 
  Our approach simplifies radically in that case.
  We are considering the tadpole diagram (\ref{tad}) with vanishing outer 
  weight for which there
  is no gluing, only a single loop gluing. Thus, the parameters
  $g$ in (\ref{Dgen}) vanish. Furthermore, 
  the initial diagram ${\cal D}_0$ may be chosen to have vanishing entries 
  only. From the structure of the basis virtual triangles and loop-gluing
  diagrams, it then follows that all the $v$ parameters vanish as well.
  The associated polytope is then characterized by the inequalities
  \ben
   0\leq l_1,...,l_r,\ \ \ \ t(l_1,...,l_r)\leq k
  \label{tr}
  \een
  The threshold level $t(l_1,...,l_r)$ is now a function of the 
  loop-gluing parameters only. In all known cases, it is a first-order 
  expression in the entries, cf. (\ref{3t}) and (\ref{k04}).
  It need not be linear, though. Assuming that this first-order
  dependence generalizes to all $su(N)$, we see that the genus-1 
0-point fusion
  multiplicity is polynomial in the level $k$, of degree less than or equal to
  the rank $r=N-1$. Let us list the lower-rank cases (see \cite{RW3}
  for $su(2)$, (\ref{sum1}) for $su(3)$, and (\ref{Nk1su4}) for $su(4)$):
  \bea
   &&r=1:\ \ \ \ N^{(k,1)}= k+1\nn
   &&r=2:\ \ \ \ N^{(k,1)}= \hf(k+1)(k+2)\nn
   &&r=3:\ \ \ \ N^{(k,1)}= \frac{1}{6}(k+1)(k+2)(k+3)
  \eea
  It is natural to expect that for general rank, the fusion multiplicity is
  \ben
   N^{(k,1)}=\bin{k+r}{r}
  \label{Ngen}
  \een
Indeed, this is the result by the Verlinde formula:
\bea
 N^{(k,h)}_{\la^{(1)},...,\la^{({\cal N})}}=\sum_{\sigma\in P_{+}^k}
  (S_{0,\sigma})^{2(1-h)}
  \biggl(\frac{S_{\la^{(1)},\sigma}}{S_{0,\sigma}}\biggr)
  \cdots\biggl(\frac{S_{\la^{({\cal N})},\sigma}}{S_{0,\sigma}}\biggr)
\eea
where $P^k_{+}$ is the set of integrable affine highest weights at level $k$.
With ${\cal N}=0$ and $h=1$, $N^{(k,1)}$ counts the number of primary 
fields at level $k$, which is exactly (\ref{Ngen}).

  \section{Conclusion}

  We have provided a prescription for characterizing higher-genus 
  ${\cal N}$-point $su(N)$ fusion multiplicities as discretized polytope 
  volumes. Our method is based on techniques of gluing BZ triangles together
  to form multi-sided diagrams. In order to treat higher-genus fusion,
  we introduced a complete basis of loop-gluing diagrams for all $su(N)$.
  The remaining input is a knowledge of threshold levels of the various
  couplings. The assignment of threshold levels to BZ triangles is
  known for $N=2,3,4$. We therefore put particular emphasis on $su(3)$ and
  $su(4)$, but also discussed the general case, assuming that the issue
  of threshold levels was settled.

  An alternative approach to fusion discussed in \cite{RW5}
  and based on \cite{Ras1,PRY,Ras2,Ras3},
  amounts to analyzing three-point functions in Wess-Zumino-Witten
  conformal field theory. Due to its universal nature, this method allows
  one to treat other Lie algebras than $su(N)$ as well. So far, 
only lower-rank
  cases have been considered explicitly.

  Related approaches to characterize fusion multiplicities by polytopes were
  considered in \cite{BCLM,paec}. 
  However, the complexity of all known methods 
  increases rapidly with the rank of the Lie algebra. It is therefore
  natural to expect that further progress will depend on
  novel insight, or an ingenious hybrid of the existing techniques.

  Finally, let us point out that all of our multiple-sum formulas for fusion
  multiplicities can be rewritten as formulas for the so-called exponential 
  sums of the corresponding polytopes.  These exponential sums are very 
  important in polytope theory \cite{Barv}, and lead to the possibility that 
  our discrete polytope
  volume picture admits a sensible description in terms of
  residue formulas (other fusion residue formulas have been written in 
  \cite{res}). We hope to report on this in the future.
  \vskip.5cm
  \noindent{\em Acknowledgments}
  \vskip.1cm
  \noindent JR thanks Frederic Lesage for helpful discussions.
The work of MAW is supported by NSERC,
  and that of JR by a CRM-ISM Postdoctoral Fellowship.

  \appendix
  \section{Higher-genus $su(4)$ fusion}

  An $su(4)$ BZ triangle is defined in terms of 18 non-negative integers:
  \ben
   \matrix{m_{14}\cr
          n_{12}~~\quad l_{34}\cr
   m_{24}~\qquad\qquad ~~m_{13}\cr
   n_{13}\qquad l_{23}\qquad n_{23} \qquad l_{24}\cr
   m_{34}\qquad\qquad\quad m_{23}\qquad\qquad\quad m_{12} \cr
   n_{14}\qquad l_{12}\qquad n_{24}\ \ \quad l_{13}\quad~~~ n_{34}\qquad
   l_{14} \cr  }
  \label{trifour}
  \een
  related to the Dynkin labels by
  \bea
   m_{14}+n_{12}=\lambda_1\ \ \ \ \ n_{14}+l_{12}=\mu_1\ \ \ \ \
   l_{14}+m_{12}=\nu_1\nn
   m_{24}+n_{13}=\lambda_2\ \ \ \ \ n_{24}+l_{13}=\mu_2\ \ \ \ \
   l_{24}+m_{13}=\nu_2\nn
   m_{34}+n_{14}=\lambda_3\ \ \ \ \ n_{34}+l_{14}=\mu_3\ \ \ \ \
   l_{34}+m_{14}=\nu_3
  \label{outfour}
  \eea
  The $su(4)$ BZ triangle contains three hexagons with the associated 
  constraints
  \ben
   \begin{array}{llll}
   &n_{12}+m_{24} =m_{13}+n_{23}\ \ \ \  
   & n_{13}+l_{23} =l_{12}+n_{24}\ \ \ \
   & l_{24}+n_{23} =l_{13}+n_{34}\nn
   &n_{12}+l_{34}=l_{23}+n_{23}\ \ \ \
   & n_{13}+m_{34} =n_{24}+m_{23}\ \ \ \
   & n_{23}+m_{23} =m_{12}+n_{34}\nn
   &m_{24}+l_{23} =l_{34}+m_{13}\ \ \ \
   & m_{34}+l_{12} = l_{23}+m_{23}\ \ \ \
   &l_{13}+m_{23} =l_{24}+m_{12}
  \end{array}
  \label{hexfour}
  \een
  Only 6 of these 9 hexagon identities are independent.  We can assign the 
  threshold level to (\ref{trifour}) as follows:
  \bea
   t&=&\max\{\la_1+\la_2+\la_3+l_{14},\ \mu_1+\mu_2+\mu_3+m_{14},\
    \nu_1+\nu_2+\nu_3+n_{14},\nn
   &&\ \ \ \ \ \ \ \la_1+\la_2+l_{14}+l_{24}+n_{14},\ 
     \la_2+\la_3+l_{14}+l_{13}+m_{14},\nn 
   &&\ \ \ \ \ \ \   \mu_1+\mu_2+m_{14}+m_{24}+l_{14},\ 
     \mu_2+\mu_3+m_{14}+m_{13}+n_{14},\nn  
   &&\ \ \ \ \ \ \   \nu_1+\nu_2+n_{14}+n_{24}+m_{14},\  
    \nu_2+\nu_3+n_{14}+n_{13}+l_{14},\nn
   &&\ \ \ \ \ \ \ l_{14}+m_{14}+n_{14}+[\hf(\la_2+\mu_2+\nu_2+l_{23}
      +m_{23}+n_{23}+1)]\}
  \label{k04}
  \eea
  The discretized volume of the convex polytope (in $v_1$, $v_2$ and $v_3$)
  subject to the inequalities
  \bea
   0&\leq&v_2,\ \mu_1-v_2,\ \la_3-v_2,\ -v_2+v_3,\ v_1-v_2,\ 
     \mu_2+v_2-v_3,\ \la_2-v_1+v_2,\nn
   &&\la_3+v_1-v_2-v_3,\ \mu_1-v_1-v_2+v_3,\ 
     n_1-v_3,\ n_2-v_1+v_2-v_3,\ n_3-v_1,\nn
   &&N_1+v_3,\ N_1'-v_3,\ N_2+v_1-v_3,\ N_2'-v_1+v_3,\ N_3-v_1,\ N_3'+v_1,\nn
   &&k-\la_1-\la_2-\la_3-N_1-v_3,\ k-\mu_1-\mu_2-\mu_3-N_3'-v_1,\ 
       k-\nu_1-\nu_2-\nu_3-v_2,\nn
   &&k-\la_1-\la_2-N_1-N_2-v_1-v_2,\ k-\mu_2-\mu_3-N_2'-N_3'-v_2-v_3,\nn
   &&k+\la_3-\nu_1-\nu_2-\nu_3-v_2-v_3,\ 
     k+\mu_1-\nu_1-\nu_2-\nu_3-v_1-v_2,\nn
   &&k-\nu_1-\nu_2-N_3'-v_1-v_3,\ k-\nu_2-\nu_3-N_1-v_1-v_3,\nn
   &&2k-\la^1+\la^3+\mu^1-\mu^3-\nu^1-\nu^2-\nu^3-v_1-v_2-v_3
  \label{cp4}
  \eea
  is the fusion multiplicity $N_{\la,\mu,\nu}^{(k)}$.
  We refer to \cite{RW4} for an explicit multiple-sum formula measuring
  this volume. Here, the parameters are defined as follows:
  \bea
  n_1&=&\la^3+\mu^3-\nu^1\nn
  n_2&=&\la^2+\mu^2-\nu^2\nn
  n_3&=&\la^1+\mu^1-\nu^3\nn
  N_1&=&-n_1+\mu_3\nn
  N_2&=&n_1-n_2+\mu_2\nn
  N_3&=&n_2-n_3+\mu_1\nn
  N_1'&=&\nu_1-N_1\nn
  N_2'&=&\nu_2-N_2\nn
  N_3'&=&\nu_3-N_3
  \label{definesu(4)}
  \eea
  where the dual Dynkin labels can be written as 
  $\la^1=\frac{1}{4}(3\la_1+2\la_2
  +\la_3),\ \la^2=\frac{1}{4}(2\la_2+4\la_2+2\la_3)$ and $\la^3=\frac{1}{4}
  (\la_1+2\la_2+3\la_3)$. $\mu^i$ and $\nu^i$ are defined similarly.

  Now we focus on the genus-1
  1-point $su(4)$ fusion.  The method for finding the 
  discretized volume of the associated convex polytope is the same as 
  for $su(3)$. In this case 
  we have 3 virtual triangles and 3 loop-gluing diagrams as shown in
  (\ref{gl4}). For the genus-1 1-point fusion this defines the diagram
  \ben
   {\cal D}={\cal D}_0+\sum_{i=1}^3v_i{\cal V}_i-\sum_{j=1}^3l_i{\cal L}_i
  \een
  with associated convex polytope
  \bea
   0&\leq&\la_1+v_2,\ -v_2,\ -\frac{1}{4}(\la_1-2\la_2-\la_3)-v_2+v_3,\nn
    &&k-\la_1-\la_2-\la_3+l_1-v_1,\ k-\frac{1}{4}(7\la_1+2\la_2+\la_3)
+l_1+l_2+
    l_3-v_2\nn
   0&\leq&\frac{1}{4}(\la_1+2\la_2-\la_3)+v_2-v_3,\ -\hf(\la_1-\la_3)-v_3,\ 
    \hf(\la_1+\la_3)+v_3,\nn
    &&k-\frac{1}{4}(5\la_1+2\la_2+3\la_3)+l_1+l_2+l_3-v_3,\ k-\la_1-\la_2
-\la_3
    +l_1+l_2-v_1-v_3\nn
 0&\leq&\hf(\la_1-\la_3)-v_3-l_3,\ -\frac{1}{4}(\la_1-2\la_2-\la_3)-v_3+v_1,\ 
    v_3-v_1-l_2,-v_1,\nn
    &&k-\frac{1}{4}(6\la_1+4\la_2+2\la_3)+l_1+l_2+l_3-v_1-v_2,\nn
    && k-\frac{1}{4}(6\la_1+4\la_2+2\la_3)+l_1+l_2-v_2-v_3\nn 
   0&\leq&-l_1+v_1,\ v_2-v_1-l_2,\ -\frac{1}{4}(\la_1-2\la_2-\la_3)+v_1-v_2,
\nn
   && k-\la_1-\la_2-\la_3+l_1+l_2-v_2-v_3,\nn
    &&k-\frac{1}{4}(6\la_1+4\la_2+2\la_3)+l_1+l_2+l_3-v_1-v_3\nn
   0&\leq&-v_2-l_3,\ v_1-v_2-v_3-l_3,\ v_2-v_1-v_3,\ v_3-v_1-v_2,\nn
    &&k-\la_1-\la_2-\la_3+l_1+l_2-v_1-v_3,\nn
   &&2k-\frac{1}{4}(10\la_1+8\la_2+6\la_3)-v_1-v_2-v_3+2l_1+2l_2+l_3
  \label{polysu(4)}
  \eea
  The fusion multiplicity $N_\la^{(k,1)}$ can now be expressed as a multiple 
  sum as follows:
  \bea
   N_\la^{(k,1)}&=&\sum_{v_3}\sum_{v_2}\sum_{v_1}\sum_{l_1}\sum_{l_2}
    \sum_{l_3}1
  \label{sumsu(4)}
  \eea
  The lower bounds of the summation variables are
  \bea
   v_3&\geq&-\hf(\la_1+\la_3)\nn
   v_2&\geq&\max\{-\la_1,\ -\frac{1}{4}(\la_1+2\la_2-\la_3)+v_3\}\nn
   v_1&\geq&\max\{\frac{1}{4}(\la_1-2\la_2-\la_3)+v_2,\ \frac{1}{4}(\la_1-
      2\la_2-\la_3)+v_3\}\nn
   l_1&\geq&\la_1+\la_2+\la_3+v_1-k\nn
   l_2&\geq&\max\{\la_1+\la_2+\la_3-l_1+v_1+v_2-k,\ \frac{1}{4}(6\la_1+4\la_2+
      2\la_3)-l_1+v_2+v_3-k,\nn
     &&\la_1+\la_2+\la_3-l_1+v_2+v_3-k,\ \la_1+\la_2+\la_3-l_1+v_1+v_3-k\}\nn
   l_3&\geq&\max\{\frac{1}{4}(7\la_1+2\la_2+\la_3)-l_1-l_2+v_2-k,\ \frac{1}{4}
      (5\la_1+2\la_2+3\la_3)-l_2+v_3-l_1-k,\nn
     &&\frac{1}{4}(6\la_1+4\la_2+2\la_3)-l_1-l_2+v_1+v_2-k,\nn
     && \frac{1}{4}(6\la_1
      +4\la_2+2\la_3)-l_1-l_2+v_1+v_2-k,\nn
     &&\frac{1}{4}(10\la_1+8\la_2+6\la_3)+v_1+v_2+v_3-2l_1-2l_2-2k\}
  \label{lowersu(4)}
  \eea
  The upper bounds of the summation variables are
  \bea
   v_3&\leq&-\hf(\la_1-\la_3)\nn
   v_2&\leq&\min\{0,\ -\frac{1}{4}(\la_1-2\la_2-\la_3)+v_3\}\nn
   v_1&\leq&\min\{0,\ v_2-v_3,\ v_3-v_2\}\nn
   l_1&\leq&v_1\nn
   l_2&\leq&\min\{v_3-v_1,\ v_2-v_1\}\nn
   l_3&\leq&\min\{v_1-v_2-v_3,\ -v_2,\ \hf(\la_1-\la_3)-v_3\}
  \label{uppersu(4)}
  \eea
  Specializing to $\la=0$, we work out the multi-summation and find the
  genus-1 0-point fusion multiplicity
  \ben
   N^{(k,1)}\ =\ \sum_{l_1=0}^k\ \sum_{l_2=0}^{k-l_1}\ 
\sum_{l_3=0}^{k-l_1-l_2}
    \ 1\ = \ \frac{1}{6}(k+1)(k+2)(k+3)
  \label{Nk1su4}
  \een

  To conclude this section, we will state how the genus-2
  0-point fusion multiplicity may be characterized as the discretized volume
  of a polytope. In this case, the associated diagram can be written as
  \bea
   {\cal D}={\cal D}_0+\sum_{i=1}^6v_i{\cal V}_i-\sum_{j=1}^3g_j{\cal G}_j-
    \sum_{k=1}^6l_k{\cal L}_k
  \eea
  The initial diagram ${\cal D}_0$ may be chosen to have only vanishing
  entries, and the associated convex polytope becomes
  \bea
   0&\leq&v_2-g_1,\ g_2-g_1-v_2,\ v_3-v_2+g_1-g_2,\nn
    &&k-v_3+g_3+l_1+l_2+l_3,\ k-v_1+l_1+g_1+g_3\nn
   0&\leq&v_2-v_3-g_2+g_3,\ g_2-g_3-v_3,\ v_3-g_3,\nn
    &&k-v_2+g_1+l_1+l_3,\ k-v_1-v_2+g_1+g_3+l_1+l_2\nn
   0&\leq&g_3-v_3-l_3,\ v_1-v_3,\ v_3-v_1-l_2,\nn
    &&k-v_2-v_1+g_2+l_1+l_2+l_3,\ k-v_2-v_3+g_1+g_2+l_1+l_2\nn 
   0&\leq&v_1-l_1,\ -v_1,\ v_2-v_1-l_2,\ v_1-v_2,\nn
    &&k-v_2-v_3+g_2+g_3+l_1+l_2,\ k-v_1-v_3+g_2+l_1+l_2+l_3\nn
   0&\leq&g_1-v_2-l_3,\ g_2-v_2-v_3-l_3,\ -v_1-v_2,\ -v_1-v_3,\nn
    &&k-v_1-v_3+l_1+l_2+g_1+g_3,\ 2k+g_1+g_2+g_3+2l_1+2l_2+l_3\nn
   0&\leq&v_4-g_1,\ g_2-g_1-v_4,\ v_5-v_4+g_1-g_2,\nn
    &&k-v_6+g_1+g_3+l_4,\ k-v_4+g_1+l_4+l_5+l_6\nn
   0&\leq&v_4-v_5-g_2+g_3,\ g_2-g_3-v_5,\ v_5-g_3,\nn 
    &&k-v_5+g_3+l_4+l_5+l_6,\ k-v_5-v_6+g_1+g_3+l_4+l_5\nn
   0&\leq&g_3-v_5-l_6,\ v_6-v_5,\ v_5-v_6-l_5,\nn
    &&k-v_5-v_6+g_2+l_4+l_5+l_6,\ k-v_4-v_5+l_4+l_5+g_2+g_3\nn
   0&\leq&v_6-l_4,\ -v_6,\ v_4-v_6-l_5,\ v_6-v_4, \nn
   &&k-v_4-v_5+l_4+l_5+g_1+g_2,\ k-v_4-v_6+g_2+l_4+l_5+l_6\nn
   0&\leq&g_1-v_4-l_6,\ g_2-v_4-v_5-l_6, -v_6-v_4,\ -v_6-v_5,\nn
   &&k-v_4-v_6+g_1+g_3+l_4+l_5,\ 2k-v_4+g_1+g_3+2l_4+2l_5+l_6
  \label{su4ex}
  \eea
  This is an example where measuring the discretized volume of the polytope 
  requires analyzing intersections of the polytope faces (in the terminology 
  of \cite{RW1}, there is no appropriate order of summation). That is in
  principle straightforward, but will not be carried out here.

\end{document}